# Effect of silver doping on the electrical properties of a-$Sb_2Se_3$

Sanjeev Gautam [a,*], Anup Thakur [b], S.K. Tripathi [a], Navdeep Goyal [a]

[a] *Centre of Advanced Study in Physics, Panjab University, Chandigarh, India*

[b] *U.C.o.E., Punjabi University, Patiala, Punjab, India*

**Abstract**

This paper reports the effect of Ag-doping on electrical properties of a-$Sb_2Se_3$ in the temperature range 230–340 K and frequency range 5–100 kHz. The variation of transport properties with thermal doping has been studied. Ag-doping produces two homogeneous phases in the sample, which are found to be voltage dependent in the temperature range studied and frequency dependent in lower frequency region (0.1–10 kHz). Activation energy $E_g$ and $C'$ [$= \sigma_0 \exp(\gamma/k)$, where $\gamma$, is the temperature coefficient of the band gap] calculated from dc conductivity has been found to vary from $(0.42 \pm 0.01)$ eV to $(0.26 \pm 0.01)$ eV and $(4.11 \pm 0.01) \times 10^{-5}$ to $(2.90 \pm 0.02) \times 10^{-6}\ \Omega^{-1}\ \text{cm}^{-1}$ respectively. Ag-doping can be used to make the sample useful in device applications.

## 1. Introduction

Chalcogenide glasses have large number of applications in IC technology and their use as photo-resists for sub-micron technology is an added advantage [1–3]. These glasses can be easily photo/thermal-doped with Ag and are useful as sensors [4] and voltage variable capacitors [5]. Ag doping also shows applications in superconducting [6] and magnetic materials [7].

Raman studies suggested the formation of homogeneous Ag–As phase after the photo-dissolution of Ag in As–S. Many investigators [4–14] have tried to explain the mechanism of silver doping and the structure of chalcogenide glasses, but the basic understanding is still incomplete. Although most of the reported work on dissolution of Ag in chalcogenide semiconductors is with thin films [8,9], but few investigators have reported this phenomenon on bulk glasses as well [8,10,12–14]. The paper reports the measurements of complex impedance parameters over a wide range of frequency and temperature for both doped and undoped samples of $Sb_2Se_3$. The doped sample becomes more conductive and also shows voltage dependence. Doping introduces two homogeneous phases in the sample (i.e., a-$Sb_2Se_3$ and Ag–$Sb_2Se_3$), which can be explained by using parallel RC network model. A structural model has been proposed to explain the results on the basis of modified-correlated barrier hopping (m-CBH) model for chalcogenides.

## 2. Experimental details

The chalcogenide glasses were prepared by melt quenching technique as described earlier [15]. The pellets were prepared by compressing the finely grounded powder of a-$Sb_2Se_3$ to maximum compactness to form circular pellets having approximately same dimensions (diameter 0.677 cm, thickness 0.040 cm). The doping of silver into the host material was carried out thermally i.e., both the faces of pellet were coated with silver paste (~10–20 μm) and then kept in an oven at temperature of 60 °C for 2 h. Another pellet (uncoated sample) was coated on both sides with aquadag (a commonly used conducting emulsion) for ohmic contacts [15]. The ohmic contacts were confirmed through linear I–V characteristics. All electrical measurements of real and imaginary components of impedance parameters ($Z'$ and $Z''$), and admittance parameters ($Y'$ and $Y''$) were made over a wide range of temperature (230–340 K) and frequency (5–100 kHz). Temperature is maintained by using a closed cycle refrigerator (Model F-70, Julabo), which has a range of ±100 °C and can maintain a constant temperature within ±1 °C for all measurements.

* Corresponding author.
*E-mail address:* sgautam@pu.ac.in (S. Gautam).

Measurements of impedance parameters were made on an ac impedance and C–V measurement system (Model 368, 410, EG& G, PARC, USA) [16] in the frequency range 5–100 kHz using dual two phase lock-in-amplifier (Model 5202, EG& G, PARC, USA). The impedance system also had a facility to obtain current–voltage (I–V) characteristics, using a potentiostat and galvanostat (Model 173, EG&G, PARC, USA).

## 3. Results

### 3.1. Dipolar behavior of Ag–$Sb_2Se_3$

Fig. 1 shows the Nyquist plots ($Z'$ vs $Z''$ plots) at different temperatures (230–340 K) for Ag–$Sb_2Se_3$. The semicircular plots indicate the dipole nature of the sample, which may be due to back and forth hopping of bipolarons between defects states ($D^+$ and $D^-$) present in chalcogenides [4].

The value of ‘*s*’ has been calculated using Jonscher’s criteria [17] from the angles for both pure and doped $Sb_2Se_3$ and plotted in Fig. 2. These values have been extracted for the smaller and bigger arc (Fig. 1), which corresponds to undoped (ss) and doped (sp) part of the sample respectively. The voltage dependence on the Nyquest plots at room temperature is plotted in Fig. 3 and capacitance with temperature at different frequencies is also studied (Fig. 4). Further the variation of ac conductivity with temperature at different frequencies is drawn in Fig. 5 and fitted according to the m-CBH Model [20]. The different parameters calculated are tabulated below

| | $W_m$ (eV) | $W_1$ (eV) | $W_2$ (eV) | $U_{eff}$ (eV) | $\varepsilon_r$ | $\tau_0$ (s) | $T_g$ (°C) |
|---|---|---|---|---|---|---|---|
| a-$Sb_2Se_3$ | 1.15 | 0.47 | 0.37 | 0.31 | 10 | $5 \times 10^{-12}$ | 130 |
| Ag–$Sb_2Se_3$ | 0.98 | 0.47 | 0.37 | 0.14 | 10 | $5 \times 10^{-12}$ | 130 |

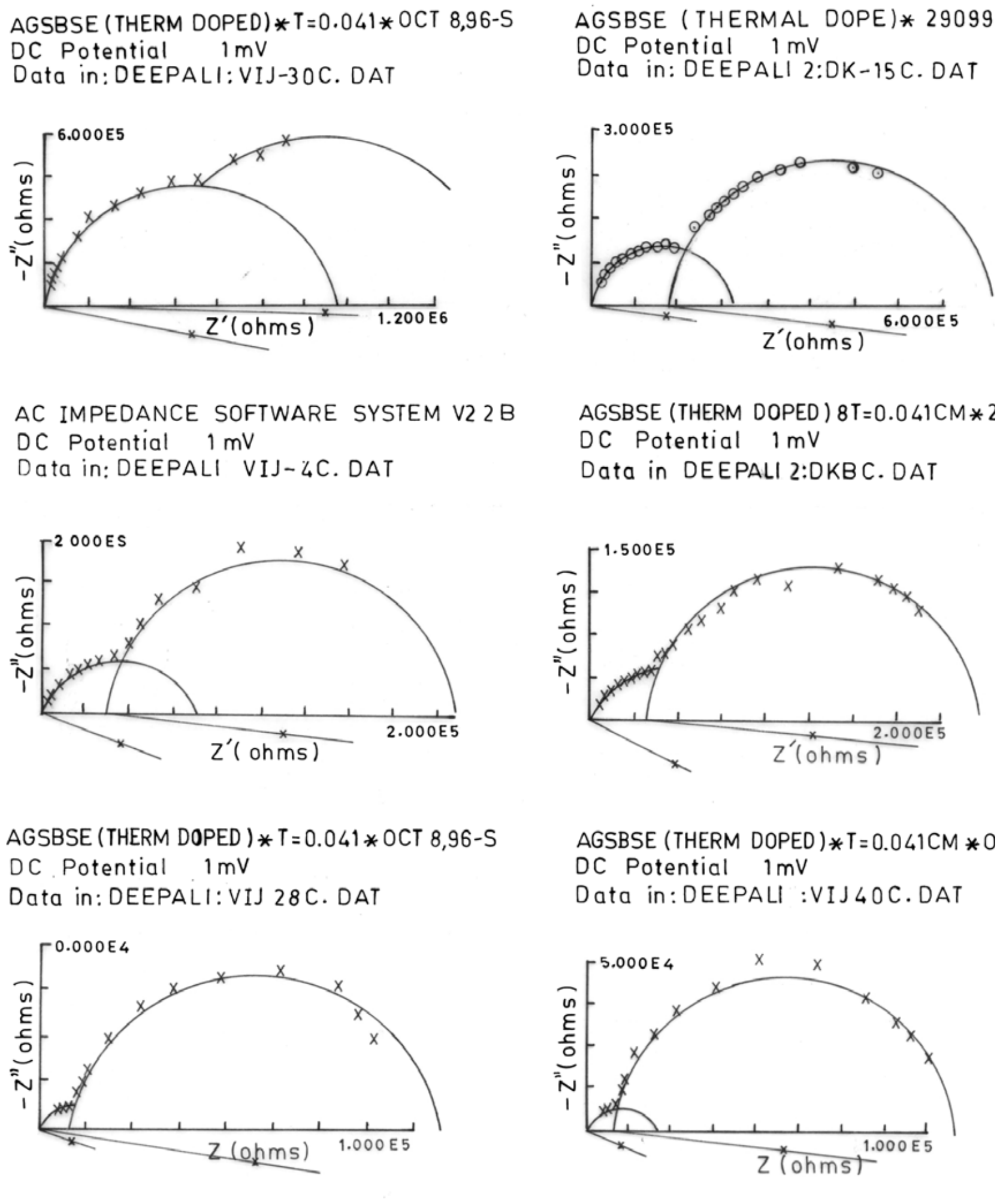

Fig. 1. Nyquist plots ($Z'$ vs $Z''$ plots) at different temperatures (243 K, 258 K, 269 K, 281 K, 301 K, 313 K) for Ag–$Sb_2Se_3$.

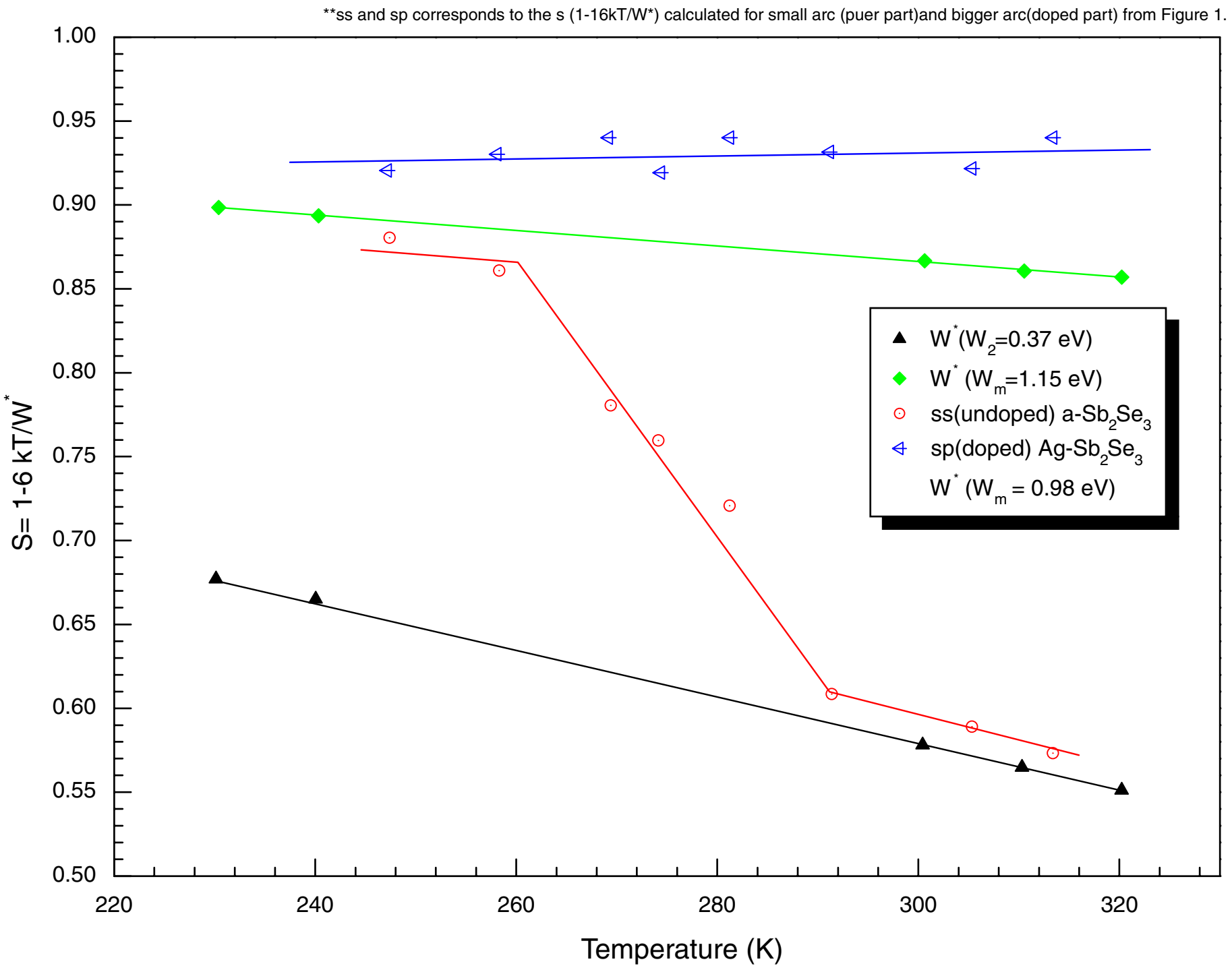

Fig. 2. The $s$ $(= 1 - 6kT/W^{*})$ values at different temperatures.

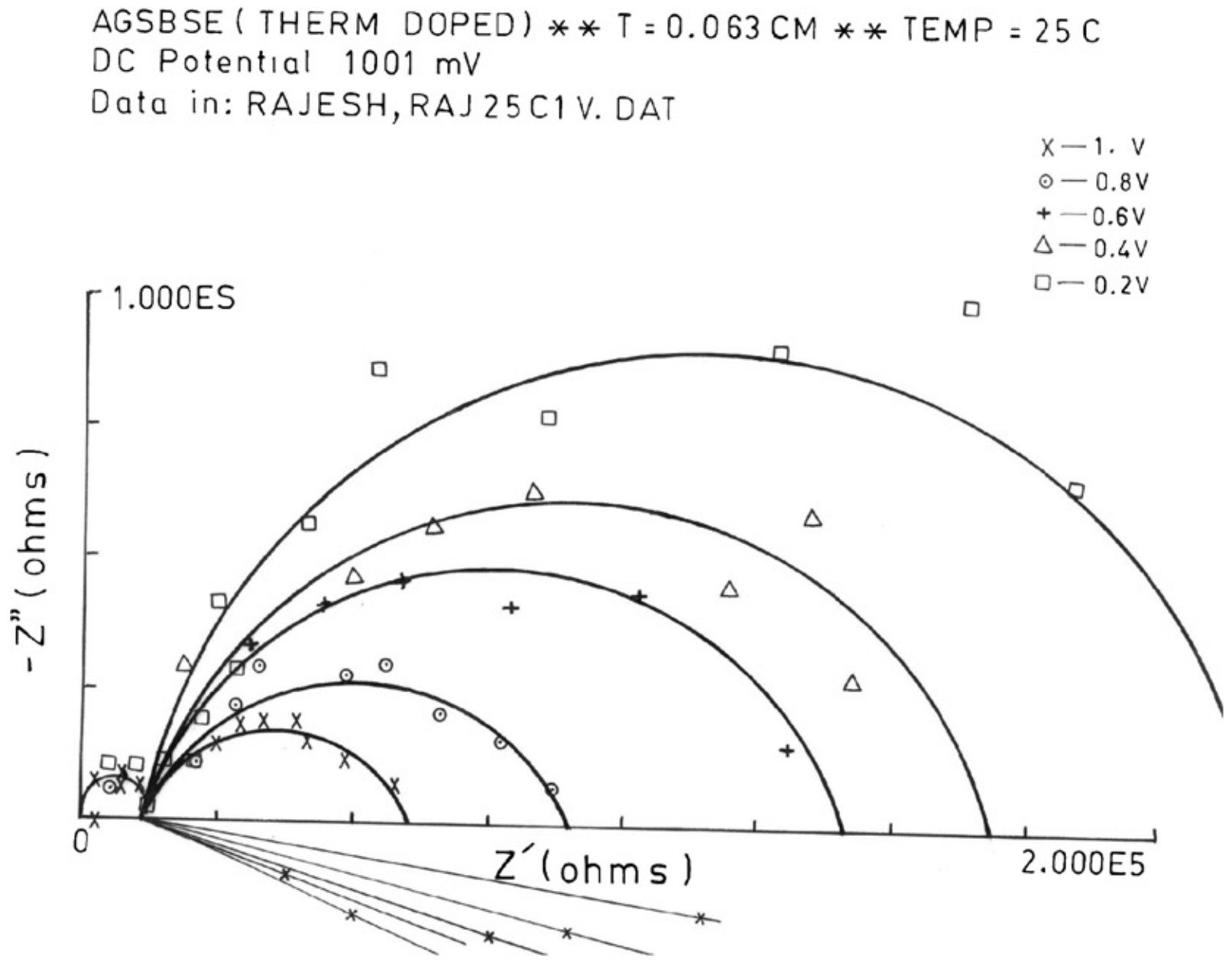

Fig. 3. Nyquist plots for Ag–$Sb_2Se_3$ at various voltage stresses.

Further, the effect of temperature and frequency on the hopping length $R_{\omega}$, which is a measure of the effective length of a dipole, is also studied using m-CBH model and drawn in Fig. 6.

### 3.2. DC conductivity

The value of dc conductivity has been obtained from I–V characteristics at different temperatures. Fig. 7(a) shows the plots of ln $\sigma_{dc}$ vs $1000/T$ for both undoped (a-$Sb_2Se_3$) and thermally doped (Ag–$Sb_2Se_3$) system. The graph obeys the pattern given by the expression

$$\sigma_{dc} = C' \exp(-\Delta E/kT), \quad (1)$$

where $\Delta E$ and $k$ are the activation energy of the material and the Boltzmann's constant respectively. At lower temperatures the sample has activation energy close to undoped material $(0.42 \pm 0.01\ \text{eV})$ while at higher temperatures (>270 K) the activation energy decreases (0.26 ± 0.01 eV) as shown in Fig 7(b). At higher temperatures the doping has distinct effect on conductivity. Similar behavior is also obtained in Ag/Cu–$As_2Se_3$ System [10].

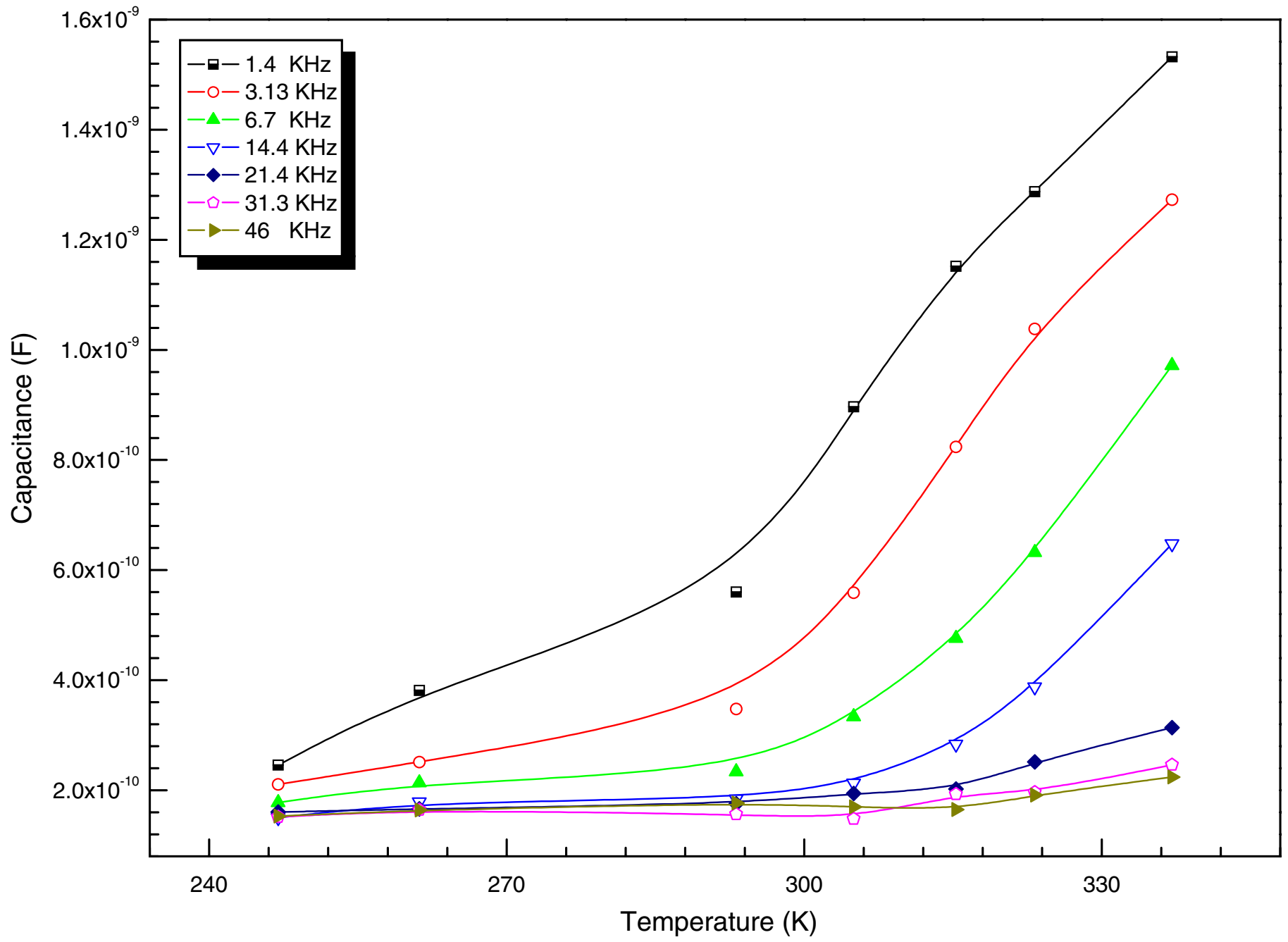

Fig. 4. Temperature variation of capacitance at different frequencies.

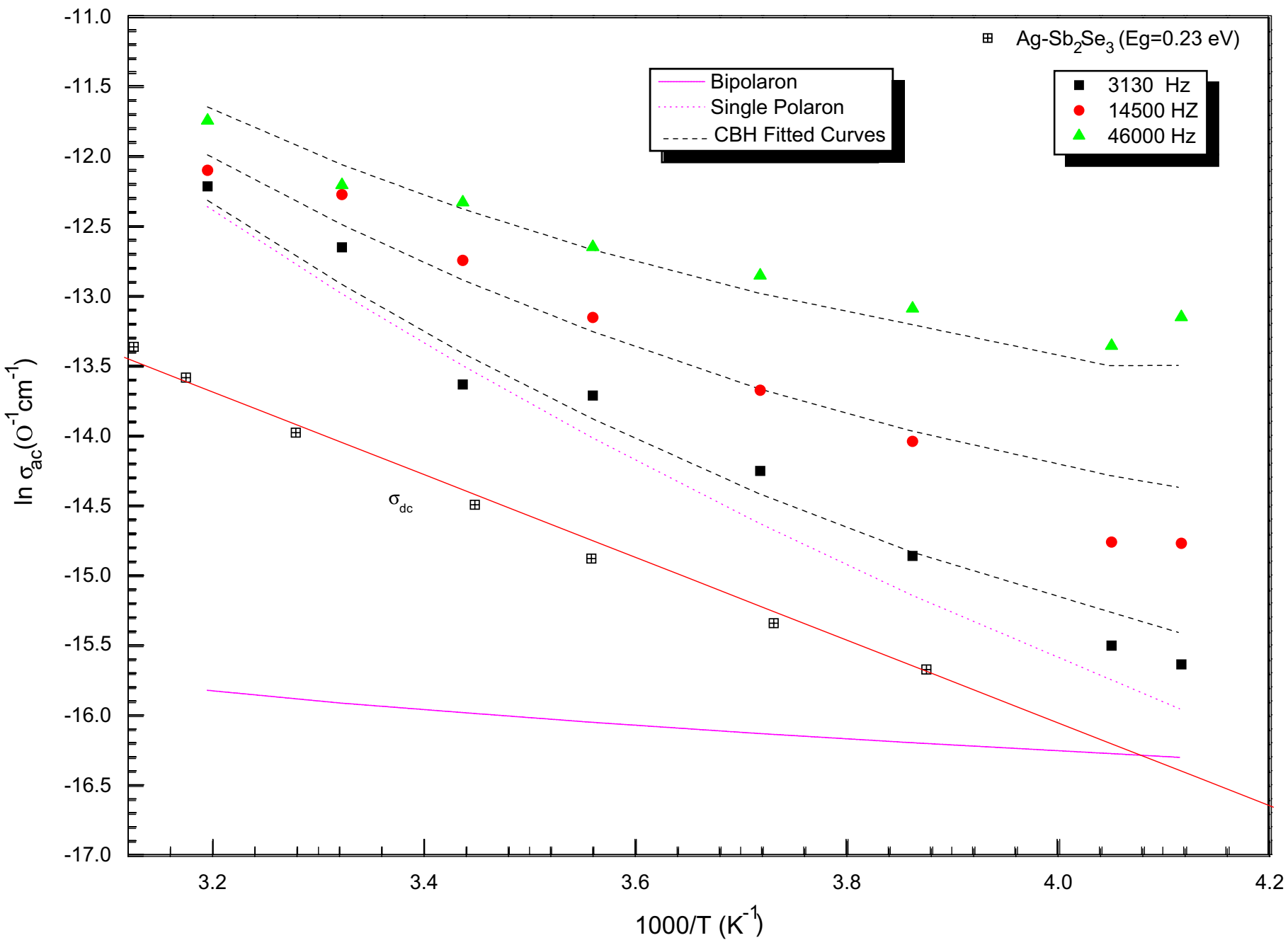

Fig. 5. Variation of $\sigma_{ac}$ ($\Omega^{-1}$ cm$^{-1}$) vs 1000/$T$ (K$^{-1}$) at different frequencies.

The value of dc conductivity calculated from Nyquist plots for a-$Sb_2Se_3$ and Ag–$Sb_2Se_3$ is also plotted in Fig. 7(b). The plot for a-$Sb_2Se_3$ is found to be linear over the temperature range studied and has single activation energy (0.42 ± 0.01 eV).

## 4. Discussion

As shown in Fig. 1, the Nyquist plots drawn are semicircular arcs with their center lying below abscissa at an angle $\alpha = (1 - s)/(\pi/2)$ (i.e., the distribution parameter $\alpha \gg 0$). Such a behavior is typical of a dielectric system involving multi-relaxing process. However for the doped system (Ag–$Sb_2Se_3$), Nyquist plots consist of two semicircular arcs, a typical behavior of two parallel R–C circuits in series as drawn in Fig. 8(a). The two semicircles are well defined with smaller arc showing a behavior similar to that of pure a-$Sb_2Se_3$, thereby indicating that the smaller arc corresponds to volume response and larger arc describes the barrier due to Ag+ [$Sb_2Se_3$] barrier (Fig. 8(b)). The real and imaginary components of impedance for such a system are given by

$$Z' = R_s + \frac{R_{p_1}}{1 + \omega^2 R_{p_1}^2 C_{p_1}^2} + \frac{R_{p_2}}{1 + \omega^2 R_{p_2}^2 C_{p_2}^2}, \quad (2)$$

and

$$Z'' = -\left[\frac{\omega C_{p_1} R_{p_1}^2}{1 + \omega^2 C_{p_1}^2 R_{p_1}^2} + \frac{\omega C_{p_2} R_{p_2}^2}{1 + \omega^2 R_{p_2}^2 C_{p_2}^2}\right]. \quad (3)$$

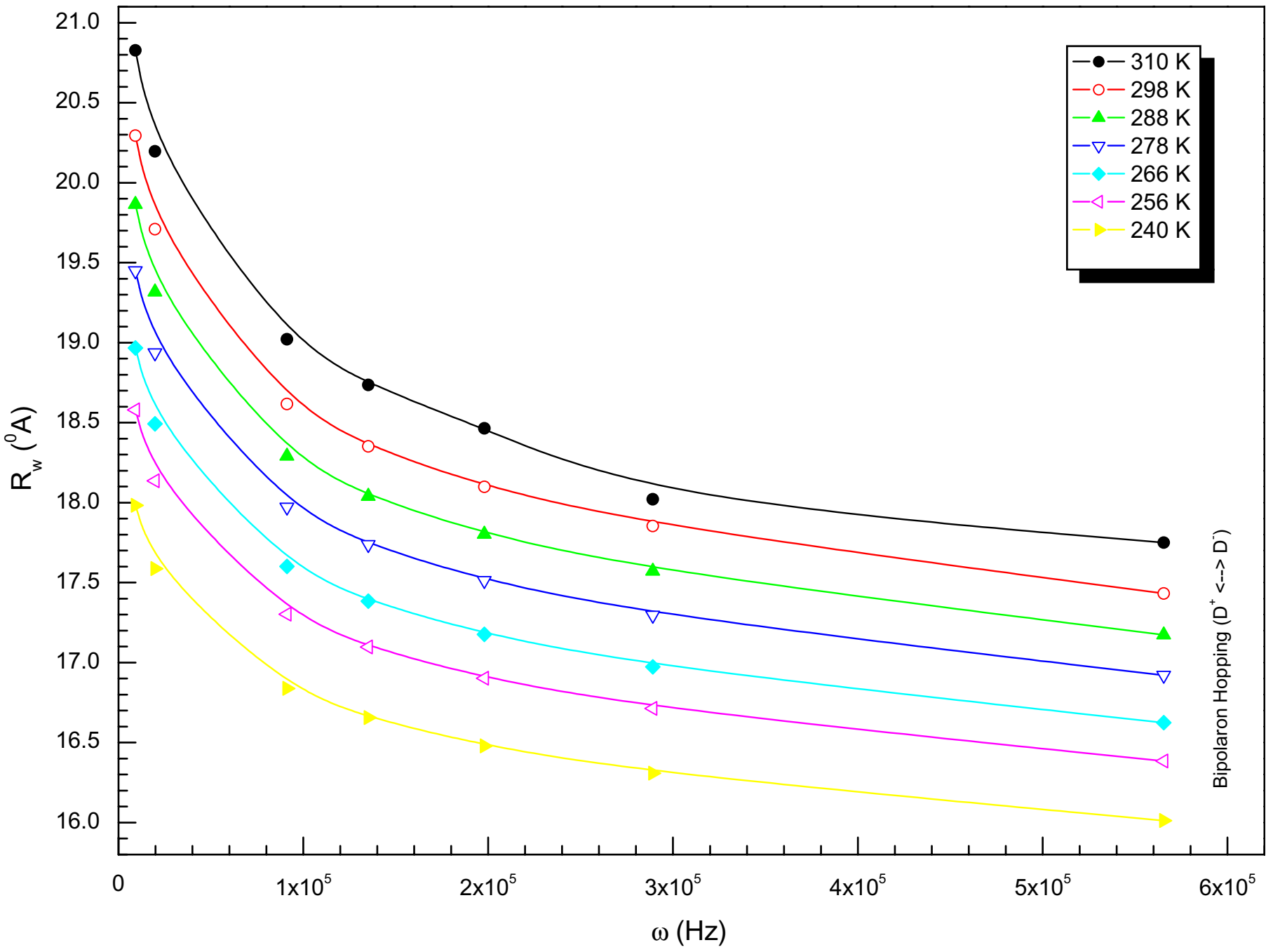

Fig. 6. Variation of $R_\omega$ (Å) with frequency ($\omega = 2\pi f$) at different temperatures.

This type of behavior represents a clear separation of time constants $G_1/C_1 \gg G_2/C_2$, with $G_1 \gg C_2$ and $G_2 \gg C_1$ [17].

Likewise, the real and imaginary components of admittance are

$$Y' = G_p/\omega \quad \text{and} \quad Y'' = \omega C_p,$$

where $G_p$ ($= 1/R_p$) and $C_p$ are the conductance and capacitance of the sample; $\omega$ is the angular frequency and $R_s$ and $R_p$ are series and parallel resistances respectively, (Fig. 8(a)), which is a characteristic feature of the sample system.

Fig. 2 indicates that the value of '*s*' decreases sharply above 270 K for a-$Sb_2Se_3$. Such a behavior has also been reported elsewhere [15]. Voltage dependence at room temperature as shown in Fig. 3 shows that the Nyquest plots reduce in the area as we increase the voltage stress and come back to its original form as it is removed. Dipolar nature of Ag–$Sb_2Se_3$ is further confirmed by the variation of capacitance ($C$) with temperature at different frequencies. It is clear from Fig. 4 that in lower temperature range the value of $C$ is nearly constant and it increases with increase in temperature. However the rate of change of capacitance (i.e., $dC/dT$) is higher for lower frequencies and decreases with increasing temperature. These results can be explained by the fact that thermally assisted hopping results in increasing the capacitance of the material. In other words, in the lower temperature range, the dipoles remain frozen and attain rotational freedom on increasing the temperature. Thus the relaxation effects are confirmed by the rate of increase in capacitance with frequency.

The complex impedance plots show the dipolar multi-relaxation nature of a-$Sb_2Se_3$. This behavior is explained by the double $R$–$C$ network model (Fig. 8(a)). However, this does not further elucidate the nature of conduction mechanism in this sample.

According to CBH model ac conductivity is given by [18,19]

$$\sigma_{ac} = \frac{n\pi^2 N N_p \varepsilon' \omega R_\omega^6}{24}, \quad (4)$$

where $n$ is the number of polarons involved in the hopping process, $NN_p$ is proportional to the square of the concentration of states and ε is the dielectric constant. $R_\omega$ is the hopping distance for conduction ($\omega\tau = 1$) and is given as

$$R_\omega = \frac{4ne^2}{\varepsilon'[W_m - kT\ln(1/\omega\tau_0)]}, \quad (5)$$

where $W_m$ is the maximum barrier height and $\tau_0$ is the characteristic relaxation time for the material.

Fig. 5 shows the variation of ac conductivity with temperature at different frequencies. It is evident from the figure that $\sigma_{ac}$ is very sensitive to temperature in the higher temperature regime. The low temperature ac conduction can be explained by considering bipolaron hopping between $D^+$ and $D^-$ centers whereas the higher temperature behavior is due to thermally activated single polaron hopping. At higher temperatures a number of thermally generated $D^0$ centers are produced with a temperature dependent concentration

$$N_0 = N_T \exp(-U_{eff}/2kT), \tag{6}$$

where $N_T$ is the concentration of $D^+$ or $D^-$ centers at $T = 0$ K. The defect concentration factor $NN_p$ in Eq. (4) is replaced by

$$NN_p = N_T^2/2 \quad \text{(for bipolaron hopping)}, \tag{7a}$$

$$NN_p = N_T^2/2\exp(-U_{eff}/2kT) \quad \text{(for single polaron hopping)}. \tag{7b}$$

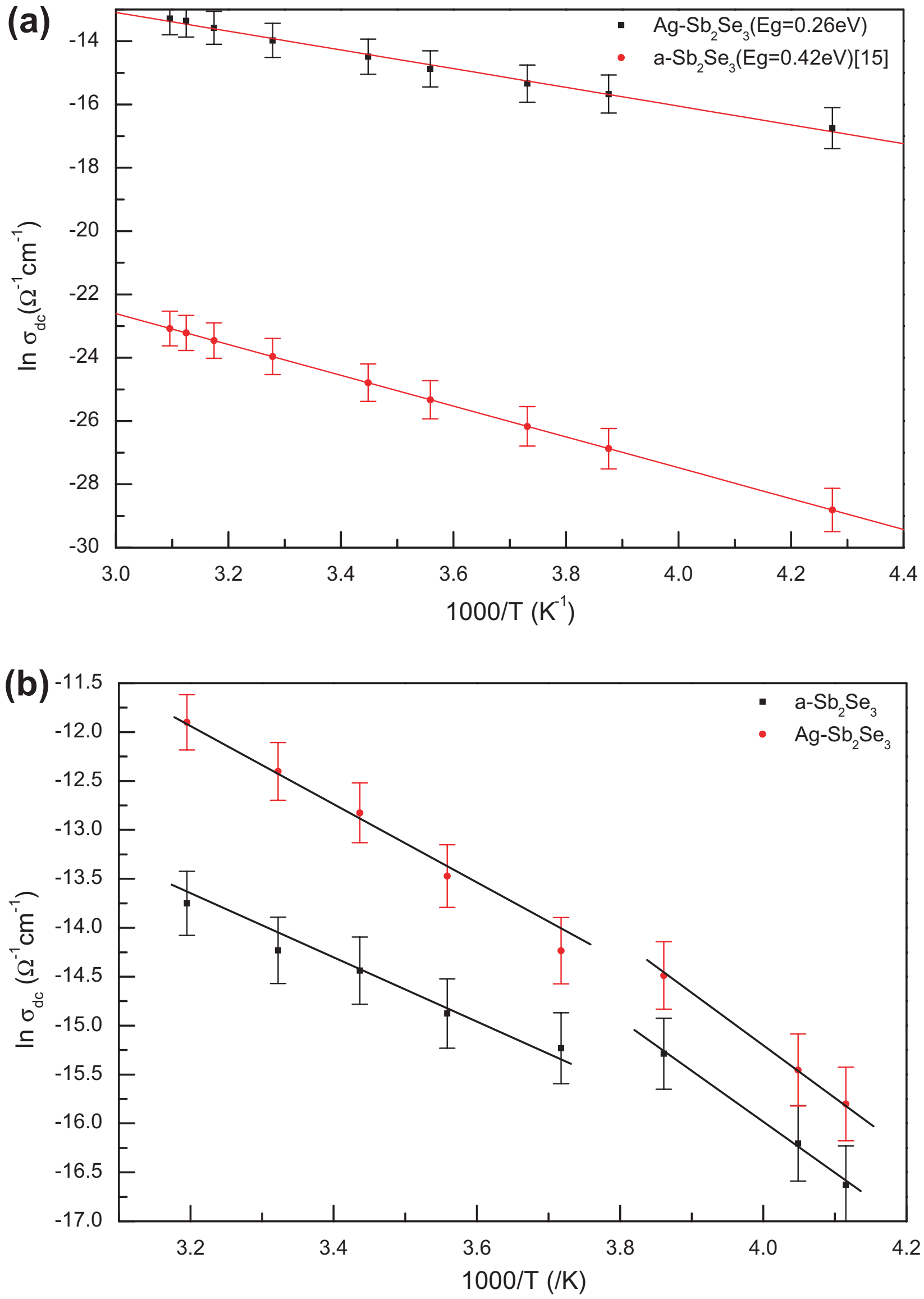

Fig. 7. Variation of $\sigma_{dc}$ ($\Omega^{-1}$ cm$^{-1}$) vs 1000/$T$ (K$^{-1}$) for a-$Sb_2Se_3$and Ag–$Sb_2Se_3$.

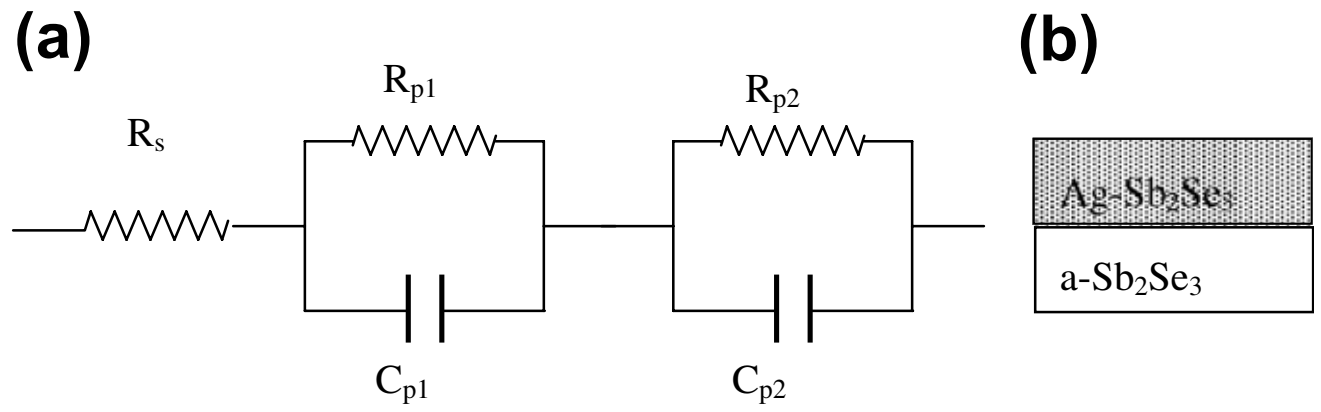

Fig. 8. *R*–*C* network model.

Thus for single bipolaron hopping this factor is thermally activated and hence $\sigma_{ac}(\omega)$ is also activated.

According to CBH model [20]

$$B = W_m - W_I + W_2, \tag{8}$$

where $B$ is optical band gap, $W_m$ is the maximum band width. The behavior with frequency $\omega$ is given by the expression

$$\sigma_{ac}(\omega) = A\omega^s. \tag{9}$$

Here [21],

$$s = \frac{\mathrm{d}(\ln \sigma_{ac})}{\mathrm{d} \ln \omega} = 1 - 6kT/W, \quad (10)$$

while $W = W_m$ for bipolaron hopping and $W = W_1$ or $W_2$ for single polaron hopping. The behavior of (Eq. (10)) ($s = 1 - 6kT/W^*$) can be seen in Fig. 2. This indicates that $W^*$ follows $W_m$ for lower temperature and $W_2$ for higher temperature for pure a-$Sb_2Se_3$, while for Ag–$Sb_2Se_3$, the curve only follows $W_m$. Fig. 2 also shows a change in conduction mechanism in a-$Sb_2Se_3$, while no such change is observed in Ag–$Sb_2Se_3$.

Fig. 6 indicates that $R_\omega$ is more sensitive to frequency in the high temperature regime and the sensitivity decreases with decreasing temperature. Thus at low temperatures, $R_\omega$ is constant with frequency, which implies a constant value of capacitance and has been found to be true from Fig. 6. This behavior is different in a-$Sb_2Se_3$ from that of Ag–$Sb_2Se_3$.

Thermal doping increases frequency dependence (Fig. 5) in the sample in the lower frequency region (0.1–10 KHz) while for higher frequency (>10 KHz) it behaves as a-$Sb_2Se_3$.

## 5. Conclusions

Dipolar nature of Ag–$Sb_2Se_3$ is verified through Nyquist plots and complex impedance studies. Activation energy of doped $Sb_2Se_3$ is different for low and high temperatures. Silver doping changes the sample into two homogeneous phases i.e., one part is doped and other part remains undoped. The doped part (Ag–$Sb_2Se_3$) becomes more sensitive to temperature and frequency and also becomes voltage dependent as compared to a-$Sb_2Se_3$ [15]. In this case bipolaron hopping dominates in the conduction mechanism, while it is both, bipolaron and single polaron in case of a-$Sb_2Se_3$.

## Acknowledgements

SG is thankful to Panjab University High Energy Physics (PUHEP) Research Group, for their support. AT is thankful to CSIR for providing financial assistance. The useful discussion with Prof. Satya Prakash is gratefully acknowledged.